\documentclass[pop,fleqn]{w-art}
\usepackage{times}
\usepackage{w-thm}

\usepackage{graphicx}
\usepackage{epstopdf}

\usepackage[dvips]{texdraw}
\usepackage{amssymb}
\def\a{\alpha}
\def\b{\beta}
\def\g{\gamma}
\def\d{\delta}
\def\e{\epsilon}
\def\ve{\varepsilon}

\def\l{\lambda}
\def\m{\mu}
\def\n{\nu}

\def\r{\rho}
\def\s{\sigma}

\def\z{\zeta}

\def\p{\psi}
\def\o{\omega}

\def\L{\Lambda}

\def\hb{\overline{h}}
\def\eq#1{(\ref{#1})}
\newcommand{\be}{\begin{equation}}
\newcommand{\ee}{\end{equation}}
\newcommand{\ba}{\begin{eqnarray}}
\newcommand{\ea}{\end{eqnarray}}
\newcommand{\ban}{\begin{eqnarray*}}
\newcommand{\ean}{\end{eqnarray*}}

\def\mW{\mathcal{W}}

\begin{document}

\DOIsuffix{theDOIsuffix}

%\Volume{51}
%\Issue{1}
%\Month{01}
%\Year{2003}

\pagespan{3}{}
%\Receiveddate{15 November 2003}
%\Reviseddate{30 November 2003}
%\Accepteddate{2 December 2003}
%\Dateposted{3 December 2003}
%%

\title[]{ N=1 domain wall solutions of massive type II supergravity and the issue of mirror symmetry}
\author[Silvia Vaul\`a]{Silvia Vaul\`a
  \footnote{Corresponding author\quad E-mail:~\textsf{silvia.vaula@uam.es}}}
\address[]{Instituto de F\'{\i}sica Te\'orica UAM/CSIC\\
Facultad de Ciencias C-XVI,  C.U.~Cantoblanco,  E-28049-Madrid, Spain}

\begin{abstract}
We report on Domain Wall solution of Calabi--Yau compactifications with general fluxes and their application to the study of mirror symmetry in generalized backgrounds. We address, in particular, to the issue of magnetic NSNS fluxes. We show that the Domain Wall gradient flow equations can be interpreted as a set of generalized Hitchin's flow equations of a manifold with  ${\rm SU}(3)\times{\rm SU}(3)$ structure  fibered along the direction transverse to the Domain wall.\end{abstract}

\maketitle

%%%%%%%%%%%%%%%%%%%%%%%%%%%%%%%%
The equivalence between type II theories compactified on Calabi--Yau (CY) mirror pairs $(Y,\,\tilde{Y})$ is a well--known property. In particular, compactification of type IIA and IIB supergravity on $Y$ and $\tilde{Y}$ respectively, gives rise to the same $N=2$ four dimensional ungauged supergravity.   The possibility to consider compactifications in the presence of $p$--form fluxes along non trivial $p$--cycles $e_I=\oint_{\g_I}\!\!F_p$ raises the question whether it is possible to generalize the notion of mirror symmetry to such more general cases.  CY compactifications of type II supergravities in the presence of fluxes yelds four dimensional gauged supergravities, the gauge parameters being proportional to the fluxes~\cite{Dall'Agata:2001zh}--\cite{D'Auria:2004tr}. It has been proved that type II supergravities  compactified on  mirror pairs $(Y,\,\tilde{Y})$ in the presence of RR fluxes, give rise to the same gauged supergravity~\cite{Louis:2002ny}. This shows that at the supergravity level type IIA and IIB RR fluxes are mapped onto one another under mirror symmetry, as can be expected from the matching of odd and even cohomology on mirror pairs. For the same reason it is unlikely  that a similar matching holds for NSNS 3--form fluxes, which reside in the odd cohomology both for type IIA and IIB.  In fact for the NSNS sector the 3--form fluxes correspond under mirror symmetry to a deformation of the CY geometry. More precisely, in the presence of  \emph{electric}  fluxes,  the $\rm{SU}(3)$ holonomy of the mirror manifold is relaxed in favor of  an $\rm{SU}(3)$ structure, the intrinsic torsion corresponding to the NSNS 3--form fluxes~\cite{FMT}.  In particular, it has been proven that CY compactification of type II supergravities in the presence of electric NSNS 3--form fluxes corresponds under mirror symmetry to  compactifications on a half--flat manifold~\cite{Gurrieri:2002wz,Gurrieri:2002iw}, where the CY cohomology is deformed according to 
\be d\a_0=-e_i\o^i;\quad d\a_a=0;\quad  d\b^A=0;\quad d\o_i=e_i\b^0;\quad d\o^i=0\,.\label{decoh}\ee
In order to fix our notations let us consider e.g. type IIB supergravity compactified on a CY $\tilde Y$ in the presence of NSNS 3--form fluxes. This corresponds to type IIA compactified on a half--flat manifold 
$\hat Y$ \eq{decoh};  in this case $A=(0,\, a)$, $a= 1,\dots h^{(2,1)} \ $ and $i=1,\dots h^{(1,1)}$.  The parameters $e_i$ correspond to the electric NSNS fluxes in type IIB compactification on $\tilde Y$, that is  $H_3=\dots- e_\L \b^\L$  with $ \L=(0,\,i)$, while the flux component $e_0$ is mapped into itself.\\
The description of mirror symmetry in the presence of \emph{magnetic} NSNS 3--form fluxes is more involved. Consider indeed the general case where $H_3=\dots +m^\L \a_\L- e_\L \b^\L$.
The choice of electric versus magnetic fluxes corresponds to a change of basis in $(\a_\L,\,\b^\L)$,  the odd cohomology of $\tilde Y$, reflecting the four dimensional electric/magnetic duality of the vector equations of motions/Bianchi identities, which holds as well in the presence of general fluxes~\cite{Sommovigo:2004vj,Sommovigo:2005fk}. Such a symplectic structure has no immediate counterpart in the even cohomology of $Y$ \eq{decoh}. Moreover, a deformation of  \eq{decoh} supporting a magnetic index would imply the presence of a closed 5--form $\chi$, such that e.g. $d\o^i\propto m^i\chi$~\cite{Benmachiche:2006df}. The following observation goes along the line of \cite{Grana:2005ny,GLW} where  it is discussed the correspondence under mirror symmetry of the magnetic NSNS fluxes and compactifications on an $\rm{SU}(3)\times\rm{SU}(3)$ structure manifold $\check Y$.\\
Indeed, introducing a symplectic basis also for the even cohomology, which includes  0-- and  6-- forms,  $(\o_\L,\o^\L)$, and indicating with $(\a_\L,\,\b^\L)$ a basis for the odd cohomology, which includes 1--, 3-- and 5-- forms,  one can write for $\check Y$
\be d\a_0=m^\L \o_\L-e_\L\o^\L;\quad d\a_a=0;\quad  d\b^A=0;\quad d\o_\L=e_\L\b^0;\quad d\o^\L=m^\L\b^0\,.\label{decoh2}\ee
Since very little is known about the properties of $\rm{SU}(3)\times\rm{SU}(3)$ structure manifolds it is not straightforward to check \eq{decoh2} by performing an explicit compactification on $\check{Y}$.  Instead, following \cite{Mayer:2004sd}, we are going to make use of Domain Wall (DW) solutions. A DW solution of the four dimensional gauged supergravity corresponding to type IIB compactification in the presence  of \emph{electric} NSNS fluxes was considered in \cite{Mayer:2004sd}.  Taking advantage of the $\rm{SO}(1,2)\times_w\mathbb{R} $ symmetry of the DW solution,  one can consider the geometry of $\hat{X}_7\equiv \hat{Y}\times_w \mathbb{R} $, where $\mathbb{R} $ parametrizes the direction transverse to the DW. If $\hat{Y}$ is a half--flat manifold then $\hat{X}_7$ is a ${\rm G}_2$ holonomy manifold, whose geometry is characterized by the Hitchin's flow equations~\cite{HitchinHF}
\be
d\,{\rm Im}\Phi_-= d\,{\rm Im}\Phi_+=0;\qquad \partial_y{\rm Im}\Phi_+=- d\,{\rm Re}\Phi_-;\qquad\partial_y{\rm Im}\Phi_-=  d\,{\rm Re}\Phi_+\label{hitch}
\ee
where $\Phi_\pm$ are the two pure spinor characterizing $\hat Y$, $d$ is the differential  along $\hat Y$ while $\partial_y$ is the derivative along the (rescaled) direction transverse to the DW.\\
We are going to generalize the result of~\cite{Mayer:2004sd} and consider~\cite{Louis:2006wq} a  DW solution of the gauged supergravity corresponding to type IIB compactification in the presence of  \emph{electric}  and \emph{magnetic} NSNS fluxes. Assuming that the geometry of  $\check{X}_7\equiv \check{Y}\times_w \mathbb{R} $ is described by Hitchin's--like flow equations~\cite{JW}, the flow equations of the DW solution provide the $y$--dependence in \eq{hitch} allowing to check  if \eq{decoh2} is suitable to describe $\check Y$.
 
Let us therefore consider the following DW metric:
\be 
g_{\m\n}(x^\m)\,dx^\mu dx^\nu = 
e^{U(z)}\hat{g}_{mn}(x^m)\,  dx^mdx^n\, - e^{-2pU(z)} dzdz \ .
\ee
where $\hat{g}_{mn}(x^m)$, $m=0,\,1,\,2$, is the metric of a three-dimensional space-time which we assume to have constant curvature and $p$ is an arbitrary real number. Using $\mu=e^{U(z)}$ instead of $z$ as the coordinate of the transverse space we arrive at 
\be\label{edef}
g_{\m\n}(x^\m)dx^\mu dx^\nu = \m^2 \hat{g}_{mn}(x^m)dx^mdx^n-\frac{d\m d\m}{\m^2
{\mW}^2(z)}\ ,
\ee
where \be\label{defmu}\mW=\pm e^{pU(z)} U^\prime(z)\ .\ee
In order to look for $\scriptstyle{1/2}$ BPS configurations we impose the following projector on the supersymmetry parameters~\cite{Mayer:2004sd,Behrndt:2001mx}
\be\ve_A=\hb A_{AB}\g_3\ve^B\label{mezzospin}\ .\ee
Here $h(z)$ is a complex function while $A_{AB}$ is a constant matrix. Consistency of \eq{mezzospin} with its hermitian conjugate implies $h\hb=1$ while $A_A^{\ B}\equiv A_{AC}\,\e^{CB}$ must be a hermitian matrix which in addition satisfies
\be A_A^{\ B}A_B^{\ C}=\d_A^C\ .\ee
Thus $A$ has to be a suitable linear combination of $(\mathbf{1},
\vec\s)$ where $\vec\s$ are the Pauli matrices.

As we are interested in the flow of the fields along the direction transverse to the Domain Wall, we assume they do not depend on the coordinates $x^m$; moreover we set to zero the vector and tensor field--strengths $F^\L_{\m\n}=H^I_{\m\n\r}=0$. With this assumption the supersymmetry transformation laws of the fermions simplify and the $\scriptstyle{1/2}$ BPS condition reads~\cite{Sommovigo:2004vj}
\ba
&&\d\p_{\m A}=D_\m\ve_A+iS_{AB}\g_\m\ve^B=0\label{psi} \\
&&\d\l^{iA}=i\partial_\m t^i\g^\m\ve^A+W^{iAB}\ve_B=0\label{la}\\
&&\d\z_\a=iP_{u A\a}\partial_\m q^u\g^\m\ve^A+N_\a^A\ve_A=0\label{zi}
\ea
where  \eq{mezzospin}  is imposed. The matrices $S_{AB}$, $W^{iAB}$ and $N_\a^A $ are the fermion shifts; they are due to the gauging and encode its properties. For the present case, and more generally for Abelian gaugings, they are related by gradient flow equations~\cite{D'Auria:2001kv}
 \be
\nabla_iS_{AB}=\frac12 g_{i\bar{\jmath}}W^{\bar{\jmath}}_{AB};\qquad
\nabla_uS_{AB}=\frac12 h_{uv}P^{v\a}_{(A}N_{B)\a}
\ee
This property gives to the conditions \eq{la} and \eq{zi} the structure of  gradient flow equations 
\be
\m\frac{dt^i}{d\m}=-g^{i\overline\jmath}\nabla_{\overline\jmath}\ln\overline W;\qquad
\m\frac{dq^u}{d\m}=- g^{uv}\partial_v\ln\overline W\label{att}
\ee
where we have used as a transverse coordinate $\m(z)=e^{U(z)}$. Equations \eq{psi}--\eq{zi} further imply that $A^{AB} S_{BC}$ is proportional to the identity or in other words
\be  \label{Wdef}
i A^{AB} S_{BC} = \frac12 W\delta^A_C \ ,\ee
where the proportionality factor defines the superpotential $W$ in \eq{att}. The condition \eq{Wdef} imposes algebraic constraints on the scalar fields which play an important role in the derivation of the solution~\cite{Louis:2006wq}.\\
Since the non--vanishing components of  the spin connection $\omega^{ab}_\m$ are given by
\be\label{oDW}
\o^{ab}_m=\hat\o^{ab}_m\ ,\qquad\o^{a3}_m=e^{(p+1)U(z)}\,U^\prime(z)
\hat{e}^a_m\ ,
\ee 
equation \eq{psi} for $\m=m$ is as well controlled by the superpotential $W$.
\be
\frac 1\ell= \frac12 e^{U} {\rm Im}(hW);\qquad U^\prime= e^{-pU}{\rm Re} (hW)\label{cosm}\ee
where $\frac1\ell$ is the cosmological constant along the DW.  From \eq{zi} we also obtain that $hW$ is real and therefore the DW is flat.  Furthermore,  we can identify
\be \mathcal{W}=\pm |W|\,.\ee   

Using \eq{att} and \eq{cosm} one can find the explicit $U(z)$ dependence of the relevant scalar fields~\cite{Louis:2006wq}.\\
Since our aim is to give a geometric interpretation in term of type IIA compactification, from now on we will interpret all the fields in this framework. In particular we have~\cite{Louis:2006wq}
\be e^{\phi_A}=C;\quad  e^{-K_V}=4\,e^{2U} ; \quad e^{-K_H}=16C^2\,e^{-6U};\quad W =\frac18
e^U(X^\L e_\L-F_\L m^\L)\ee
Note that the IIA dilaton is constant. The K\"ahler potentials of the vector--  and hyper-- multiplets are defined in terms of the moduli of $\hat Y$ in type IIA on  compactifications as
\be e^{-K_V}\equiv i\left[\bar X^\L F_\L - \bar F_\L X^\L\right];\qquad  e^{-K_H}\equiv i\left[\bar Z^A W_A - \bar W_A Z^A\right]\label{Kpot}\ee
where $\L=0,\dots n_V$ and $A=0,\dots n_H$ being $n_V$ and $n_H$ the number of  vector--  and hyper-- multiplets, respectively. They parametrize the  K\"ahler class and the complex structure deformations $n_V=h^{(1,1)}$, $n_H=h^{(1,2)}$.\\
The complex scalars in \eq{la} are the special coordinates of the vector multiplets special K\"ahler manifold $t^i=\frac{X^i}{X^0} $. The scalars $z^a=\frac{Z^a}{Z^0}$ span the special K\"ahler submanifold inside the quaternionic manifold spanned by the $q^u$ \eq{zi}. 

In order to describe the geometry of the internal manifold $\check Y$, we can conveniently rewrite equations \eq{att} and \eq{cosm}. Before we perform a change of coordinates $\partial_z=e^{(p+3)U}\partial_w$ and define a rescaled section
\be\label{Zrescale} (Z^A,\,W_A)=|c|\, (Z^A,\,W_A)_\eta\ , \qquad |c|^2\equiv
e^{K_V-K_H}\ee
We can now rewrite \eq{att} as  
\be\partial_w\begin{pmatrix}{{\rm Im}X^\L}\cr {{\rm Im }
F_\L}\end{pmatrix}=-\begin{pmatrix}{m^\L}\cr {e_\L}\end{pmatrix}\label{dRephi+}\ee
\be\partial_w\begin{pmatrix}{{\rm Im}Z^A}\cr{ {\rm Im }W_a}\cr{ {\rm Im
}W_0 }\end{pmatrix}_{\!\!\!\!\eta}=-|c|\begin{pmatrix}{0}\cr{ 0}\cr{ {\rm Re}X^\L e_\L- {\rm Re}F_\L
m^\L }\end{pmatrix}\ .\label{dRephi-}\ee
and the zero cosmological constat condition \eq{cosm} as
\be{\rm Im}X^\L e_\L- {\rm Im}F_\L m^\L=0\ .\label{dImphi+}\ee
Let us now define two pure spinors according to the symplectic invariant basis \eq{decoh2}
\be
\Phi_+=X^\L\o_\L-F_\L\,\o^\L\ ,\qquad
\Phi_-=Z^A_\eta\a_A-W_{\eta A}\,\b^A\ .\label{p2}
\ee 
The derivative along the direction transverse to the DW is easily computed using \eq{dRephi+}, \eq{dRephi-}
\be
\partial_y{\rm Im}\Phi_+=e_\L\,\omega^\L-m^\L\, \omega_\L;\qquad
\partial_y{\rm Im}\Phi_-= -|c|({\rm Re}X^\L e_\L- {\rm Re}F_\L
m^\L)\,\beta^0\ . \label{gen1}
\ee
Using  \eq{decoh2} we obtain
\be
d\,\Phi_+ =(X^\L e_\L -F_\L m^\L)\, \beta^0;\qquad
d\,\Phi_- =  |c|^{-1} ( m^\L\o_\L-e_\L\o^\L)\label{gen2}
\ee
Up to a change of coordinates $dy= |c|^{-1} dw$, equations  \eq{gen1}, \eq{gen2} are equivalent to \eq{hitch}, where now the pure spinors $\Phi_\pm$ are defined in \eq{p2}.\\
Finally, let us discuss the properties of the seven-dimensional manifold $\check{X}_7$. As the metric on the DW is flat  and the background  $M_{(1,2)}\times_w \check{X}_7$ solves the string equation of motion, we expect $\check{X}_7$ to be Ricci flat. For half-flat manifolds this was indeed shown in  refs.\ \cite{Mayer:2004sd,HitchinHF,CS}. In order to discuss the
generalization at hand let us introduce 
the seven dimensional exterior derivative by
\be \hat d=d+dy\,\partial_{y}\, ,\ee  
where $d$ acts on $\check{Y}_6$ and $\partial_{y}$ is the derivative with respect to the coordinate of $\mathbb{R}$.  Furthermore, following \cite{JW} one can define the generalized forms $\rho$ and $*\r$
on $\check{X_7}$ which are given in terms of $\Phi_\pm$ by
\be
\rho=-{\rm Re}\Phi_+\wedge dy-{\rm Im}\Phi_-\ , \qquad
*\r={\rm Re}\Phi_- \wedge dy+ {\rm Im}\Phi_+\ .\label{startre}
\ee 
$*\rho$ is the Hodge dual of $\rho$  with respect to the generalized metric. As noted in \cite{JW} the equations \eq{gen1}--\eq{gen2} then correspond to
\be d\r=*d\!*\r=0\ ,\ee
and imply that $\check{X_7}$ has an integrable $G_2\times G_2$ structure and is indeed Ricci-flat.

\begin{acknowledgement}
This work has been supported  by the Spanish Ministry of Science 
and Education grant BFM2003-01090, the Comunidad de Madrid grant HEPHACOS 
P-ESP-00346 and by the EU Research Training Network {\em Constituents, 
Fundamental Forces and Symmetries of the Universe} MRTN-CT-2004-005104
\end{acknowledgement}

\end{document}